\documentclass[pre,preprint,amsmath,showpacs,amssymb]{revtex4}
\usepackage{graphicx,psfrag}
\begin{document}
\title{ Phase synchronization in an array of driven Josephson junctions}
\author{Chitra R N }
\email{rchitra@cusat.ac.in}
\author{V C Kuriakose}
\email{vck@cusat.ac.in} %
\affiliation{Department of Physics, Cochin University of Science
and
Technology, Kochi, 682022} %
\date{\today}
\begin{abstract}
We consider an array of N Josephson junctions  connected in parallel
and explore the condition for chaotic synchronization. It is found
that the outer junctions can be synchronized while they remain
uncorrelated to the inner ones  when  an  external  biasing is
applied. The stability of the solution is found
out for the outer junctions in the synchronization manifold.
Symmetry considerations lead  to  a  situation  wherein the inner
junctions can synchronize for certain values of parameter. In the
presence of a phase difference between  the  applied  fields, all
the junctions exhibit phase synchronization. It  is also found that
chaotic motion changes to periodic in the presence of phase
differences.
\end{abstract}

\pacs{05.45.Xt, 05.45.Gg}

 \maketitle

\textbf{Due to the application of chaotic synchronization in
secure communication to brain modeling a great deal of
investigation has been done in this field. The presence of even
a small phase difference between the applied fields was found to
desynchronize a completely synchronized system. Also the phase
difference was found to have application in taming chaos in
dynamical systems. Recently it was observed that the end lasers
in an array of three laser system was found to synchronize while
it remained uncorrelated with the middle laser which originally
connected the two.  In this work we study an array of Josephson
junctions in the presence a phase difference between the driving
fields and its effect  on synchronization and suppression of
chaos.}

\section{{i}ntroduction}
Chaos in Josephson junction (JJ) has been studied extensively after
its presence was demonstrated using numerical simulation
\cite{huber}. When we treat JJs within the Stewart-McCumber model,
the equation   describing  the  behaviour  of JJ is identical to the
equation for a driven damped pendulum which has been studied
theoretically for several routes to chaos\cite{humier,braiman}. Thus
JJ becomes an ideal physical system  to study chaos. The  rf- biased
JJs find practical importance in  the construction of devices like
parametric amplifiers, voltage  standards, pulse generators, SQUID
for  detection  of  very  weak magnetic fields, etc.
\cite{fesser,schieve,pozzo}. {For  these devices,  it  is essential
to  avoid all  types  of  noise,  chaos etc. JJs consisting of
Superconductor-Insulator-Normal metal-Insulator-Superconductor
(SINIS) showing non-hysteretic I-V characteristics with high damping
has been fabricated for programmable dc-voltage standards
\cite{schul} or ac-voltage standards based on synthesis of
calculable wave forms\cite{benz}}.

Pecora and Carroll in 1990 reported  that synchronization of
chaotic systems \cite{pecora}  could  be  achieved, since  then
different types of synchronization such as complete, generalized
and phase synchronization of chaotic oscillators have been
described theoretically and observed experimentally
\cite{pik,kur}. Synchronized  chaotic  oscillations  have  been
found  in  many nonlinear systems  like  lasers,  neural
network,  etc\cite{terry,shua}. Chaotic synchronization also
find application in communication. It was demonstrated using
R\"{o}ssler oscillators that during the transmission of
information about a stimulus through an active array, the
stimulus created the way to be transmitted by making the chaotic
elements to phase synchronize \cite{bapt}. The stability of
synchronous state is analyzed by Lyapunov function method
\cite{hal} and the master stability approach \cite{pec}. Phase
difference between the applied fields plays an important role in
suppressing chaos and the synchronization of chaotic systems.
Duffing oscillator was studied for the effect of phase
difference on chaotic synchronization \cite{yin}. Josephson
junction has been investigated for both periodic and chaotic
synchronization. Coupling between self generated Josephson
oscillations through a microwave transmission line was found to
play an important role in collective synchronization of JJ array
\cite{kim}.  In a system of two JJs in parallel, the phase
difference between the applied fields was found to bring chaotic
motion to a periodic one for a large range of parameter values
\cite{chitra}. A parallel array of coupled short JJs linked
together by inductors has been used to fabricate highly
sensitive detectors \cite{chevr}. Although there is extensive
work  on synchronization  of  coupled  JJs, studies  on chaotic
synchronization  of  JJs  is  much  less.

In this work we analyze a parallel array of  N-coupled JJs with
parameters lying in the chaotic regime and study synchronization
of the system. The  paper  is  organized  as follows. In section
 II we discuss the model for an array of JJs linked in parallel
with linking resistor $R_s$ in between. Section  III  contains
the study of  the synchronization in such an array and discuss
the stability of the synchronous solution. The effect of phase
difference between the applied fields on synchronization and its
role in suppressing chaos is also discussed. Results  are
summarized in section IV.

\begin{figure}[tbh]
\centering
\includegraphics[width=10cm]{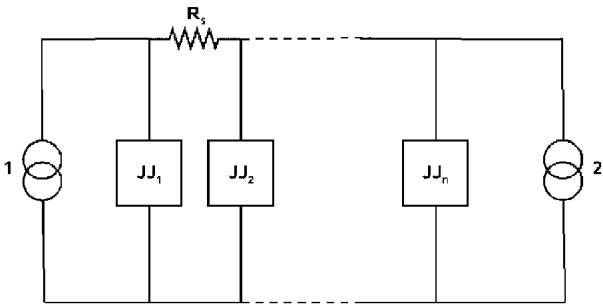}%
\caption{Schematic representation of an array of JJ linked in parallel with a linking resistor $R_s$.
1 and 2 are the driving fields.
} %
\label{figone}
\end{figure}

\section{The Model}
The equation of a single Josephson junction represented by the
resistively and capacitively shunted junction (RCSJ) model can
be written by solving Kirchoff's law as
\begin{equation}
\frac {\hbar C} {2e}\frac {d^2 \phi} {dt'^2} + \frac {\hbar}
{2eR}\frac {d \phi} {dt'}+ i_{c}\sin\phi=
i'_{dc}+i_0'\cos(\omega t')
 \label{single}
\end{equation}
where $\phi$ is the phase difference of the wave function across
the junction, $i'_0\cos(\omega t')$ is the driving rf - field and
$i'_{dc}$ is the dc bias. The junction is characterized by a
critical current $i_c$, capacitance $C$ and normal resistance
$R$. The coupled JJ considered here consists of a pair of such
junctions wired in parallel with a linking resistor $R_s$
\cite{blackburn}. A schematic representation
of an array of JJ wired in parallel with linking resistors is
given in fig \ref{figone}. The equation of motion for  an array of  N
coupled  current driven JJs can be written in the normalized
form as
\begin{subequations}
\label{equationone1}
\begin{eqnarray}
\label{equationone} \ddot{\phi_1}+ \beta \dot{\phi_1}+
\sin\phi_1&=&i_{dc}+i_0 \cos(\Omega t)
-\alpha_s\left[\dot{\phi_1}-\dot{\phi_2}\right] \\\nonumber
\vdots\hspace{1.5cm}  &\vdots&     \hspace{1.5cm} \vdots    \\
\ddot{\phi_i}+ \beta \dot{\phi_i}+ \sin\phi_i&=&
\alpha_s\left[\dot{\phi}_{i+1}+\dot{\phi}_{i-1}-2 \dot{\phi_i}\right] \\\nonumber
\vdots\hspace{1.5cm}  &\vdots&     \hspace{1.5cm} \vdots    \\
\ddot{\phi_N}+ \beta \dot{\phi_N}+
\sin\phi_N&=&i_{dc}+i_0\cos(\Omega t)
-\alpha_s\left[\dot{\phi_N}-\dot{\phi}_{N-1}\right]
\end{eqnarray}
\end{subequations}
where $i$ varies from 2 to N-1 and the dimensionless damping parameter
$\beta$ is defined as
$$\beta=\frac{1}{R}\sqrt{\frac{\hbar}{2 e i_c}}.$$   The normalized time scale is written as $t=\omega_{J1} t'$
where $\omega_{J1}=\left(2ei_{c1}/\hbar
C_{1}\right)^{\frac{1}{2}}$. The dc bias current $i_{dc}'$ and
the rf amplitude $i_{0}'$ are normalized to the critical current
$i_{c1}$. The actual frequency $\omega$ is re-scaled to
$\Omega=\omega /\omega_{J1}$ and the coupling factor is defined
as $\alpha_{s}=\left(R_1 / R_s\right)\beta$.

\begin{figure}[tbh]
\centering
\includegraphics[width=8cm]{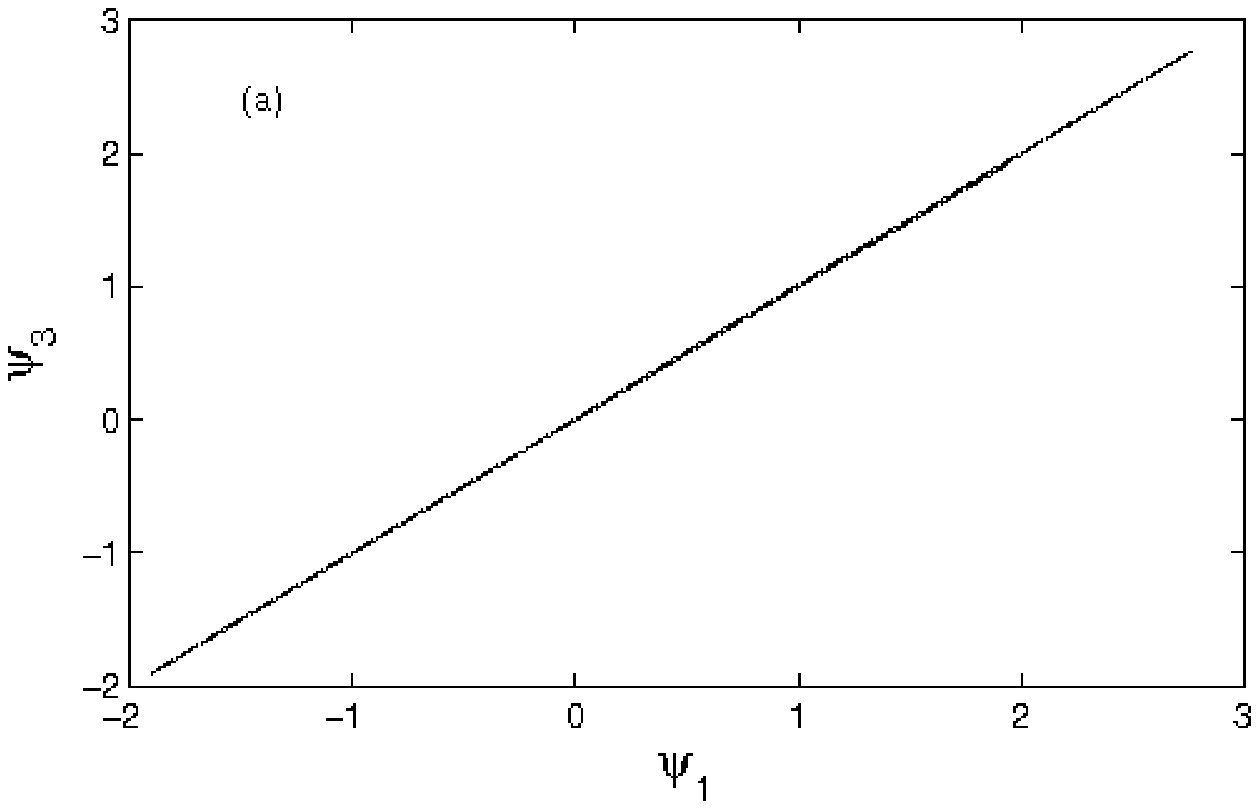}%
\hspace{1in}%
\includegraphics[width=8cm]{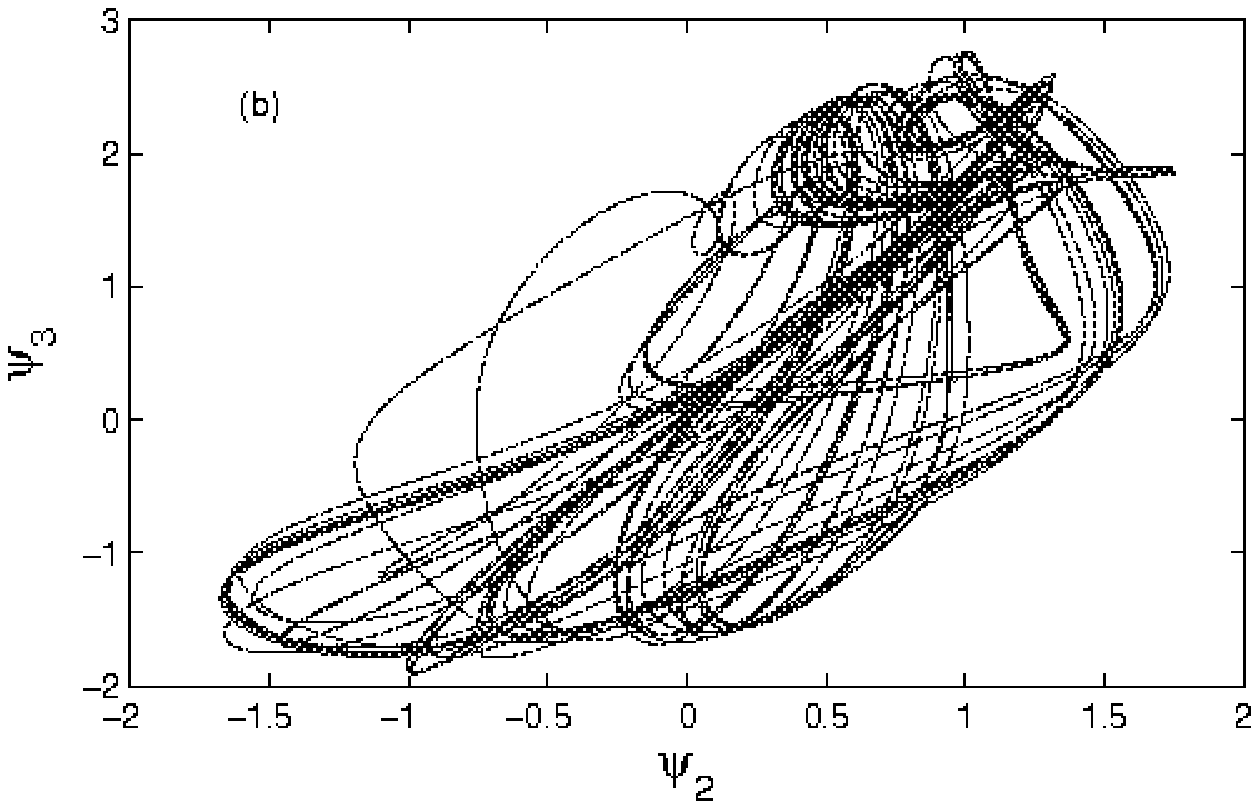}
\caption{(a) The outer juntions are synchronized (b) Outer
junction and middle junction is uncorrelated. The parameter
values are
$\beta=0.3,i_0=1.2,\omega=0.6,i_{dc}=0.3,\alpha_s=0.37.$
} %
\label{syncthree}
\end{figure}

The Josephson junction is  chaotic for the parameter values
$\beta=0.3, i_0=1.2, \omega=0.6$ and $i_{dc}=0.3$. We fix these
parameter values for the numerical simulations. The junctions
are taken to be identical and for a coupling strength of
$\alpha_s=0.37$, the outer junctions synchronize while the inner
junction remain uncorrelated with the two outer ones. It can be
seen from Fig \ref{syncthree}(a) that the outer junctions are
synchronized whereas Fig.\ref{syncthree}(b) shows that it is
uncorrelated with the middle junction for an array of three JJs.

\section{Stability analysis}
In order to perform  the stability analysis for the synchronized
state  of  N-coupled Josephson junctions, we first consider three
JJs linked in parallel. In the first order form the three identical
junctions can be written as
\begin{subequations}
\label{equationtwo} %
\begin{eqnarray}
\label{twosubone} %
\dot{\phi_1} &=& \psi_1\\\nonumber %
\dot{\psi_1}&=& -\beta \psi_1-\sin\phi_1+i_{dc}+i_0
\cos(\Omega t)
-\alpha_s \left[\psi_1-\psi_2\right]\\
\label{twosubtwo} %
\dot{\phi_2} &=& \psi_2\\\nonumber %
\dot{\psi_2}&=& -\beta \psi_2-\sin\phi_2+
\alpha_s \left[\psi_1+\psi_3-2 \psi_2\right]\\
\label{twosubthree} %
\dot{\phi_3} &=& \psi_3\\\nonumber %
\dot{\psi_3}&=&-\beta \psi_3-\sin\phi_3+i_{dc}+i_0 \cos(\Omega
t+\theta)-\alpha_s \left[\psi_3-\psi_2\right]
\end{eqnarray}
\end{subequations}
From eq.\ref{twosubone} and \ref{twosubthree} it can be observed
that the outer junctions are identical and symmetric with
interchange of variables in the absence of a phase difference
$\theta$ between the applied fields. Hence there exists an
identical solution  for the outer systems given by
$\phi_1=\phi_3=\phi(t)$ and this type of behavior where systems
show identical behavior is called complete synchronization. Due
to asymmetry  the middle junction may have different dynamics.
The stability of the synchronous solution of the outer junctions
is analyzed by two methods.

We define the difference variables
$\phi_{13}^-=\frac{\phi_1-\phi_3}{2}$ and
$\psi_{13}^-=\frac{\psi_1-\psi_3}{2}$ and the approximate  dynamics
transverse to the synchronization  manifold is obtained by linearizing the
corresponding subsystem consisting of the outer junctions. The equation  may be given as
\begin{eqnarray}
\label{equation3}
\dot{\phi}_{13}^-&=& \psi_{13}^-\\ \nonumber %
\dot{\psi}_{13}^-&=& -\beta \psi_{13}^-- \cos\phi_{13}^+\sin\phi_{13}^--\alpha_s\psi_{13}^-
\end{eqnarray}
Linearizing eq. \ref{equation3} we get the approximate dynamics
transverse to the synchronization manifold. In terms of the
Jacobian matrix we can rewrite the above equation as
\[ \left( \begin{array}{c}\dot{\phi}_{1,3}^-\\ \dot{\psi}_{1,3}^-\\ \end{array} \right) =
\left( \begin{array}{cc}
0 & 1 \\
\cos\phi_1 & -\beta-\alpha_s \\
 \end{array} \right) \left( \begin{array}{c}\phi_{1,3}^-\\ \psi_{1,3}^-\\ \end{array} \right),\]\
where $\sin\phi_{1,3}^-\approx\phi_{1,3}^-$ and
$\cos\phi_{1,3}^+\approx\cos\phi_1$ as $\phi_1\approx\phi_3$ in
the synchronization manifold. The eigen values of the matrix are
\begin{equation}
\label{eigen}
m_{1,2}=-\frac{(\alpha_s+\beta)}{2}\left[1\pm\sqrt{1+\frac{4
\cos\phi_1}{(\alpha_s+\beta)^2}}\right]
\end{equation}
The stability of the synchronous state is controlled by the
eigen values $m_{1,2}$ \cite{vincent}. If $m_{1,2}$ are complex
conjugates with negative real part, the corresponding solution
is stable. In the above case the average of the term in the
radical is found and it is a complex number with real part
greater than unity. The real part of the largest eigen value is
thus found to be negative and hence satisfy the criterion for
stability of synchronization.

As a second test, we follow the method given by Landsman et.al
\cite{alex} where the conditional Lyapunov exponents are
calculated with respect to the perturbation out of the
synchronization manifold.  Eq. \ref{equationtwo} reduces to a
set of four equations in the synchronized state as the outer
junctions may be represented by a single set of equations. In
terms of the synchronous solutions $\phi(t)$ and $\psi(t)$, we
can define variables  $\Delta\phi(t)=\phi_1(t)-\phi(t)$ and
$\Delta\psi(t)=\psi_1(t)-\psi(t).$ Linearizing transverse to the
synchronization manifold, we have
\begin{equation}
\label{jacob1} \frac{d\Delta\phi_i}{dt}=J\Delta\phi_i
\end{equation}
and
\begin{equation}
\label{jacob2} \frac{d\Delta\psi_i}{dt}=J\Delta\psi_i
\end{equation}
where i=1,3 and $J$
is the Jacobian matrix evaluated at $\Delta\phi(t)$ and
$\Delta\psi(t).$ Thus we have
\[ \left( \begin{array}{c}\Delta\dot{\phi}_{1,3}\\ \Delta\dot{\psi}_{1,3}\\ \end{array} \right) =
\left( \begin{array}{cc}
0 & 1 \\
1 & -\beta-\alpha_s \\
 \end{array} \right) \left( \begin{array}{c}\Delta\phi_{1,3}\\ \Delta\psi_{1,3}\\ \end{array} \right)\]\
$ \Delta\phi_{1,3}$ and $\Delta\psi_{1,3}$ are the perturbations of the outer oscillators from the synchronous solution
 $\left\lbrace \phi(t),\psi(t) \right\rbrace.$

The Wronskian of the linearized system can be related to the
trace of the matrix by the Abel's formula \cite{alex}
 \[ W(t) = \left| \begin{array}{cc}
\Delta\phi & \Delta\psi  \\
\Delta\dot{\phi} & \Delta\dot{\psi}  \\
 \end{array} \right|=exp\left(\int_{0}^{t}(-\alpha_s-\beta)dt'\right)\]
where we have dropped the subscripts of the linearized variable. The Wronskian gives the phase space dynamics of the system. Taking the natural log of the Wronskian we get
\begin{equation}
\ln[W(t)]=\ln|\Delta\phi\Delta\dot{\psi}-\Delta\psi\Delta\dot{\phi}|=-\int_{0}^t(\alpha_s+\beta)
dt,
\end{equation}
which is a monotonically decreasing function of time.
 The sum of the conditional Lyapunov exponents is given as,
\begin{equation}
\label{lyapunov} \sum_{j=1}^{M}\lambda_j=\lim_{t\rightarrow
\infty}
\frac{1}{t}\ln|\det(\Phi(\Delta\phi_{1,3},\Delta\psi_{1,3})(t)|
\end{equation}
where $\Phi$ is the matrix solution of eqns. \ref{jacob1} and
\ref{jacob2}. The sum of the conditional Lyapunov exponents can
be now approximated as
\begin{equation}
\lambda_1+\lambda_2\approx-(\alpha_s+\beta).
\end{equation}
The sum of the conditional Lyapunov exponents is  negative
indicating that the phase space of the coupled system shrinks to
a trajectory representing the synchronous solution. Thus the two
methods lead to the same conclusion.

\begin{figure}[tbh] \centering
\includegraphics[width=9cm]{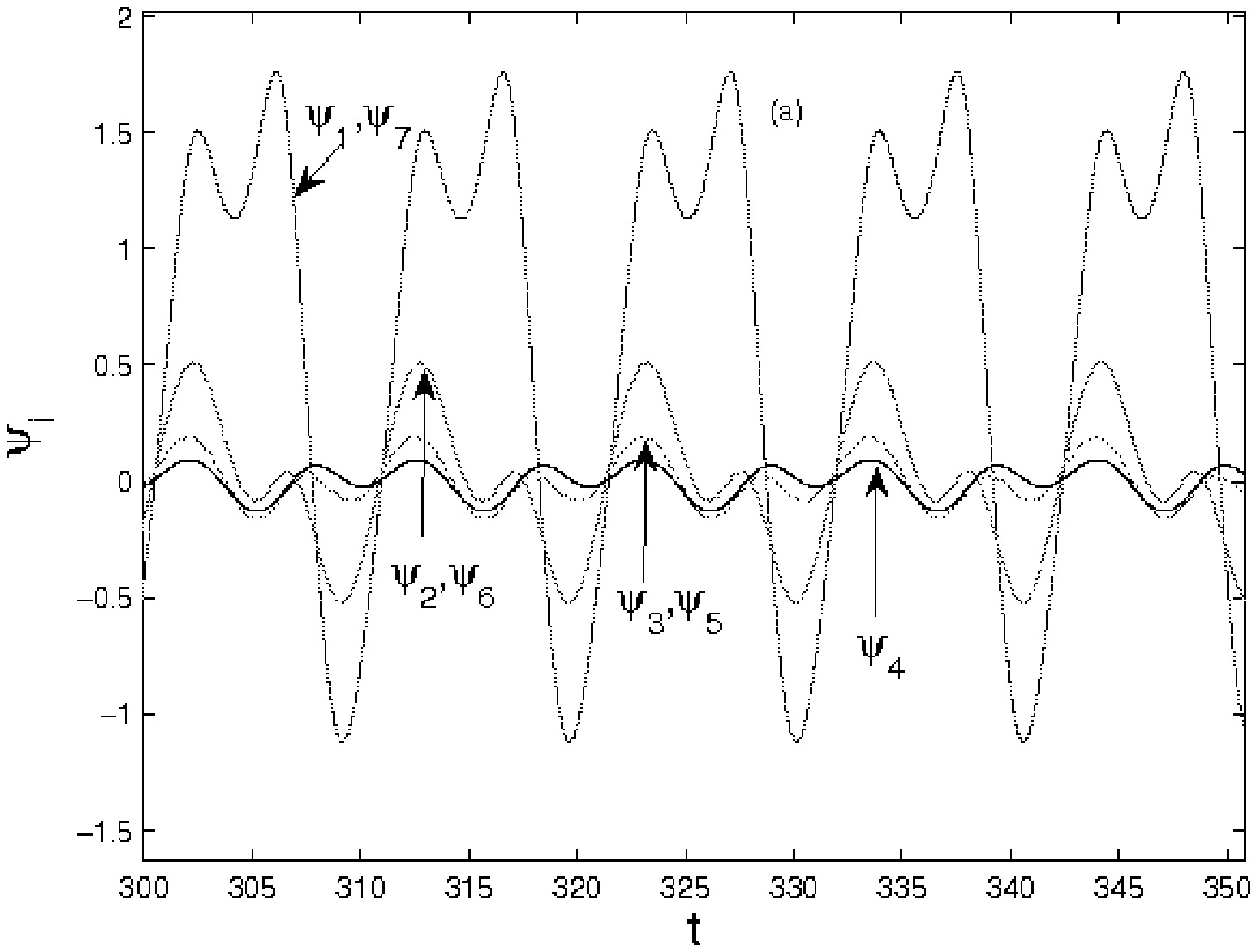}%
\hspace{1in}%
\includegraphics[width=9cm]{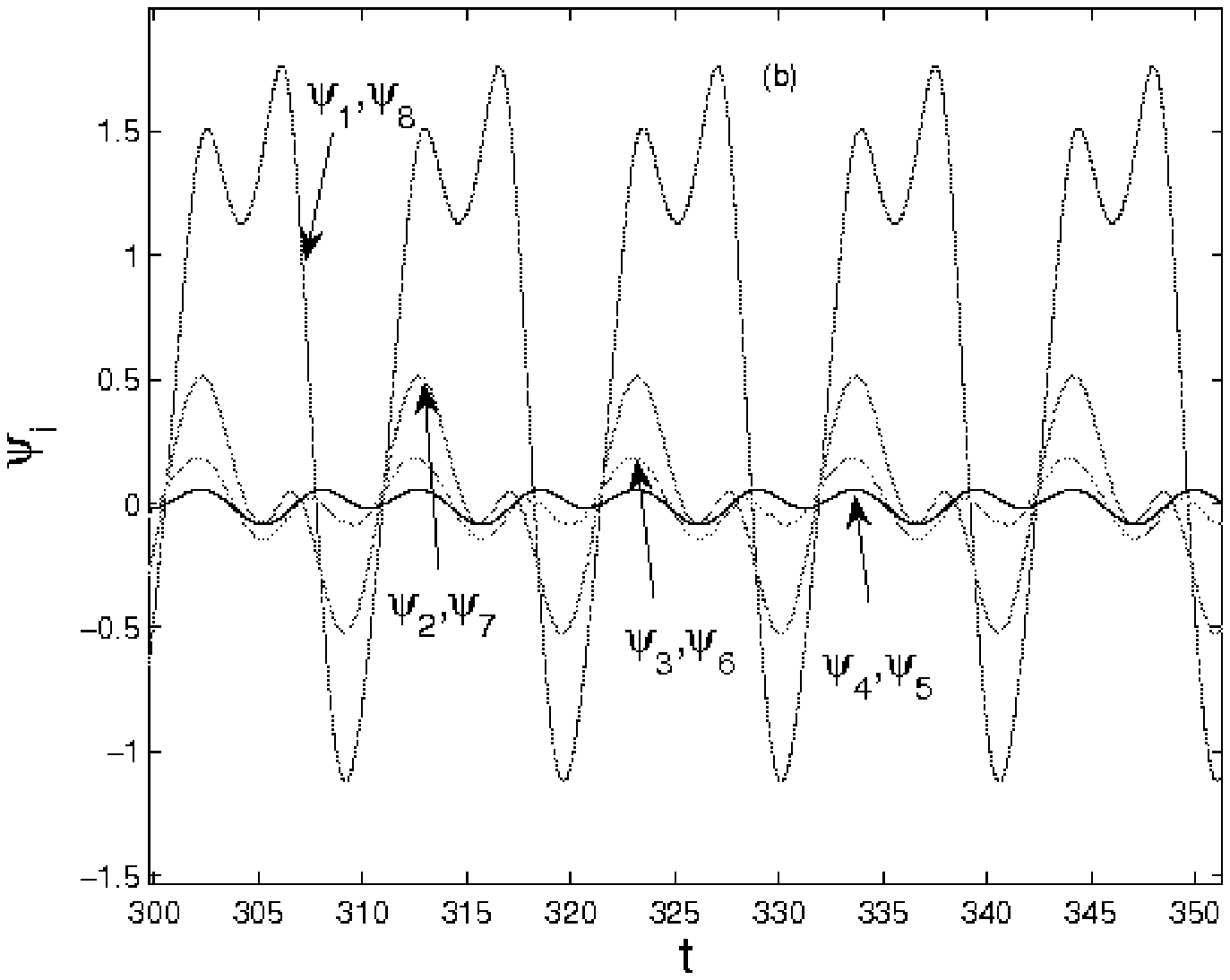}
\caption{(a)shows the time series plot for an array of $7$ JJs
and (b) for 8 junctions.
$\beta=0.3,i_0=1.2,\omega=0.6,i_{dc}=0.3,\alpha_s=0.37$and
$\theta=0.5 \pi.$
} %
\label{arraysyn}
\end{figure}

Now we analyze the subsystem constituted by the outer and the
middle junctions. We define new variables
$\phi_{i2}^-=\frac{\phi_i-\phi_2}{2}$ and
$\psi_{i3}^-=\frac{\psi_i-\psi_2}{2}$ where $i=1,3$. As the
outer junctions are identical,  it is enough to study any one
subsystem. So considering the case with $i=1$, we write,
\begin{eqnarray}
\label{rungesubaaa}
\dot{\phi}_{12}^-&=& \psi_{12}^-\\ \nonumber %
\dot{\psi}_{12}^-&=& -\beta \psi_{12}^--
\cos\phi_{12}^+\sin\phi_{12}^-+\frac{1}{2}\left[i_{dc}+i_0\cos(\Omega
t)\right]-\alpha_s(\frac{3 }{2}\psi_{12}^-).
\end{eqnarray}
From Eq. \ref{rungesubaaa} we conclude that  in the presence of
an external applied field it is not possible to synchronize all
the three junctions due to the asymmetry induced by the applied
fields. However in the absence of an external field, an
identical solution can exist for all the three junctions.
\begin{figure}[tbh]
\centering
\includegraphics[width=7cm]{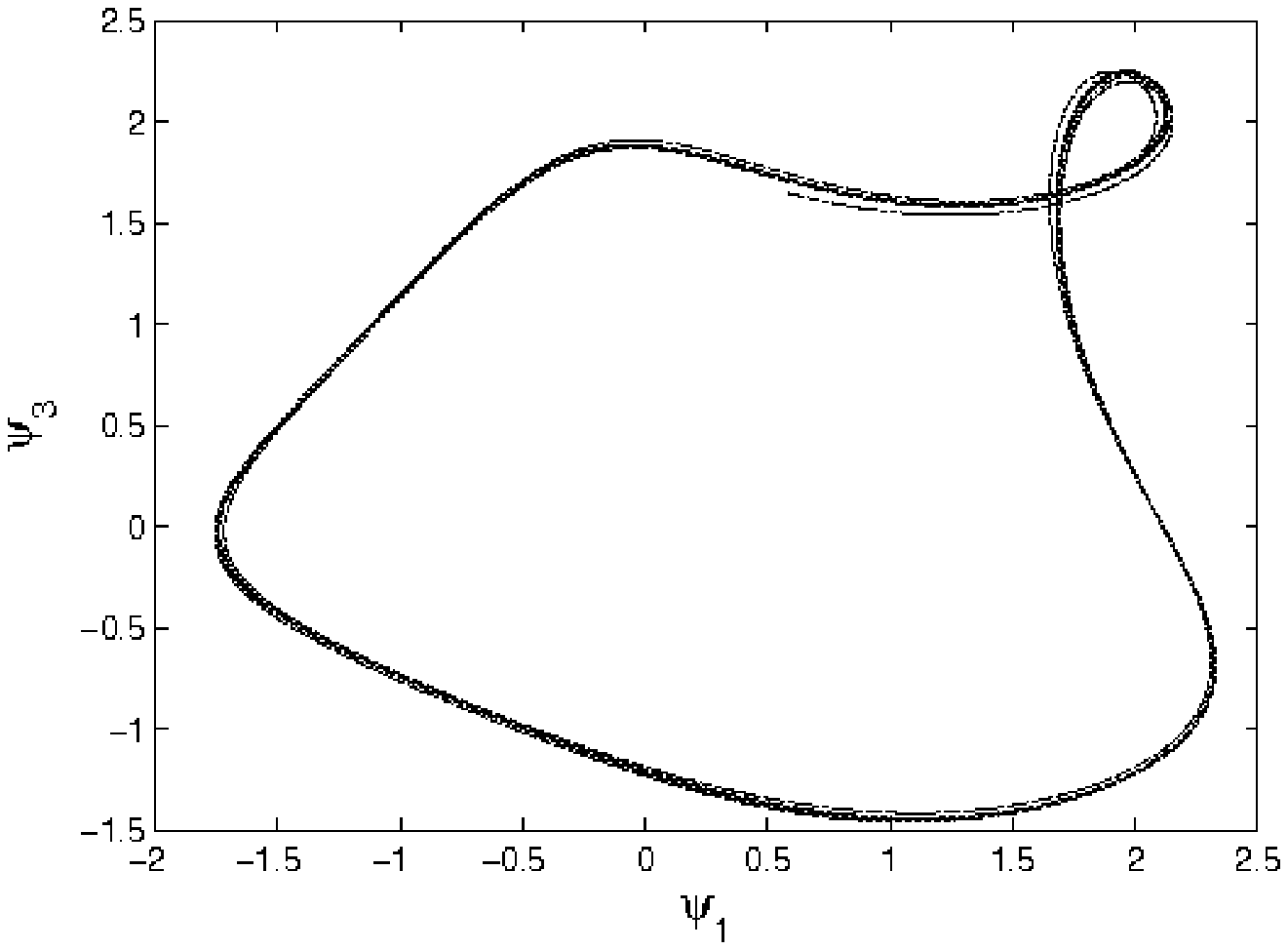}%
\hspace{1in}%
\includegraphics[width=7cm]{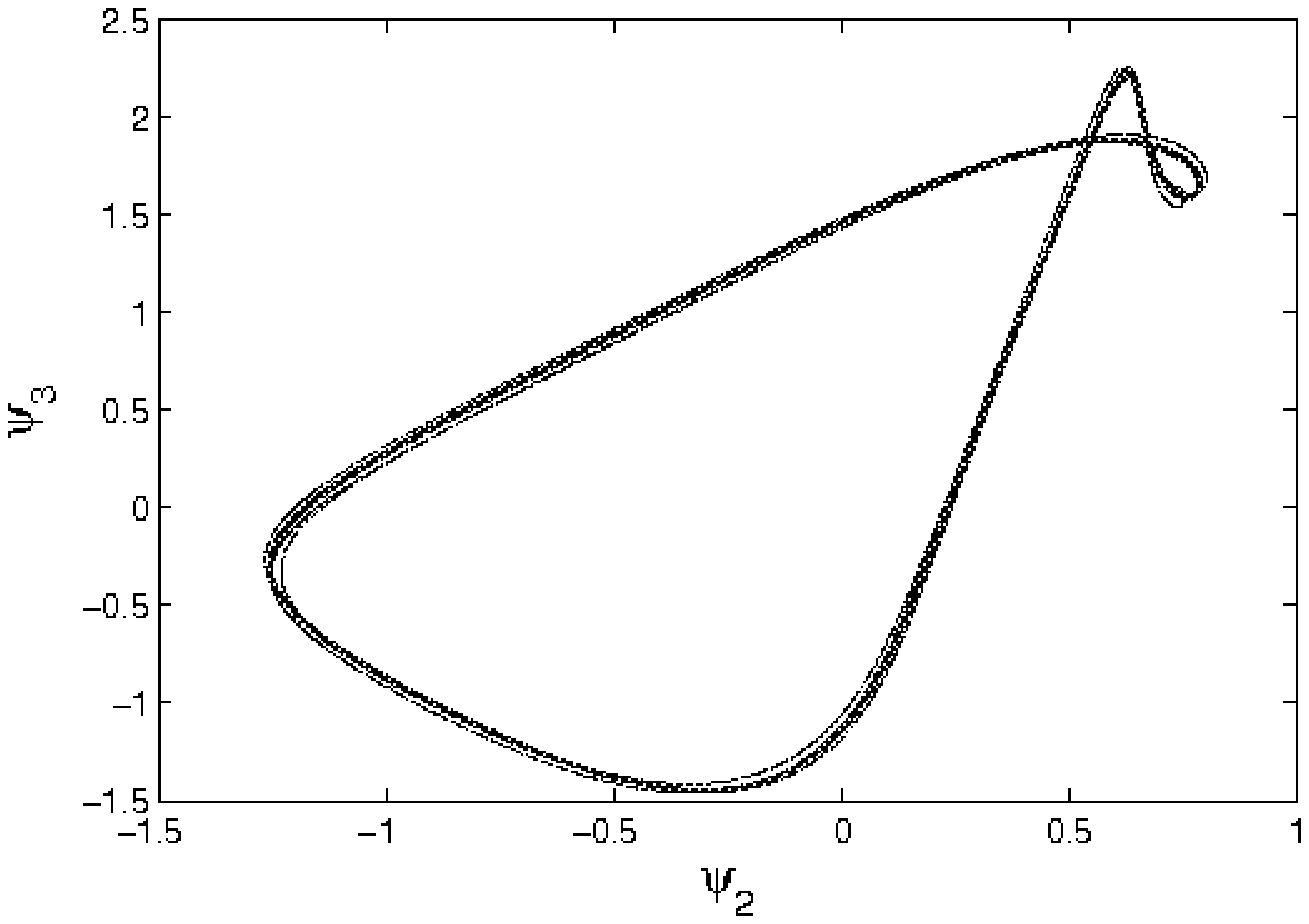}
\caption{(a) and (b) shows that the junctions are phase
correlated.
$\beta=0.3,i_0=1.2,\omega=0.6,i_{dc}=0.3,\alpha_s=0.37$and
$\theta=0.5 \pi.$
} %
\label{powtau7}
\end{figure}
Extending the symmetry analysis to a system  of N JJs coupled in
parallell with nearest neighbour coupling, the second and the
$(N-1)^{th}$ junction may have an  identical solution for
certain parameter values. Similarly, the third and the
$(N-2)^{nd}$ junctions may have identical solutions and so on.
Thus in the case of an array, from symmetry considerations we
may deduce that $N/2$ solutions may exist if there are even
number of junctions in the array and $\frac{N+1}{2}$ solutions
will be present for odd number of junctions. The time series
plot for an array of $7$ and $8$ junction is plotted in
Fig.\ref{arraysyn}. It can be observed from
Fig.\ref{arraysyn}(a) that in an array of seven JJs the four
solutions exists for the parameter range considered. The fourth
junction has an independent solution. In Fig.\ref{arraysyn}(b)
we have plotted the time series for $8$ JJs.

The presence of  a phase difference between the applied fields
changes the scenario completely. On the application of a small
phase difference between the applied fields, the outer junctions
desynchronize and all the three junctions are thus uncorrelated.
But for sufficiently large values of phase differences, all the
three junctions are found to be in phase synchronization.
Considering the difference variables
$\psi_{1,2},\psi_{1,3},\psi_{3,2}$ as defined earlier, we
explain the phenomena as follows. Due to the asymmetry that
arises between the outer junctions in the presence of the phase
difference we need the extra variable $\psi_{3,2}$ to analyze
this situation. The equations for the three difference variables
may be written by substituting eq.\ref{equationtwo}  as
\begin{subequations}
\label{rungephase} %
\begin{eqnarray}
\label{rungepha} %
\dot{\psi}_{12}^-&=& -\beta \psi_{12}^-- \cos\phi_{12}^+\sin\phi_{12}^-+\frac{1}{2}\left[i_{dc}+i_0\cos(\Omega t)\right]-\alpha_s(\frac{\psi_{12}^-}{2}+\psi_{32}^-)\\
\label{rungepha1} %
\dot{\psi}_{13}^-&=& -\beta \psi_{13}^-- \cos\phi_{13}^+\sin\phi_{13}^- + i'_0\sin(\Omega t+\frac{\theta}{2})-\alpha_s(\psi_{13}^--\psi_{32}^-)\\
\label{rungepha2} %
\dot{\psi}_{32}^-&=& -\beta \psi_{32}^--
\cos\phi_{32}^+\sin\phi_{32}^-+\frac{1}{2}\left[i_{dc}+i_0\cos(\Omega
t+\theta)\right]-\alpha_s(\frac{\psi_{32}^-}{2}+\psi_{12}^-),
\end{eqnarray}
\end{subequations}
where $i'_0=i_0\sin\frac{\theta}{2}.$ Thus each subsystems
experiences a different driving field with the same frequency
but different phases. Due to the phase relationship between the
driving fields, a definite phase relationship is found to exist
between all three junctions.

\begin{figure}[tbh]
\centering
\includegraphics[width=6cm]{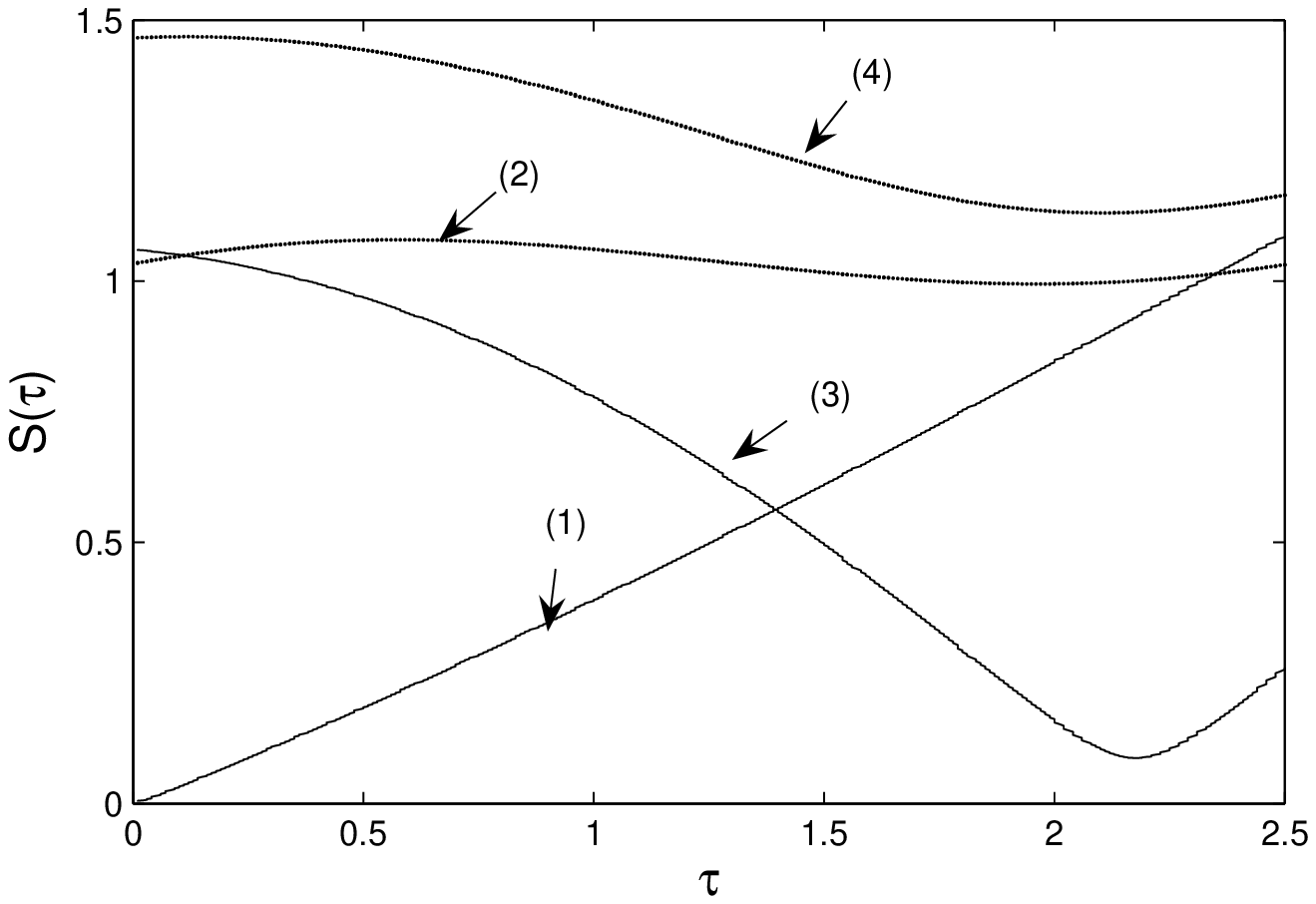}%
\hspace{1in}%
\includegraphics[width=6cm]{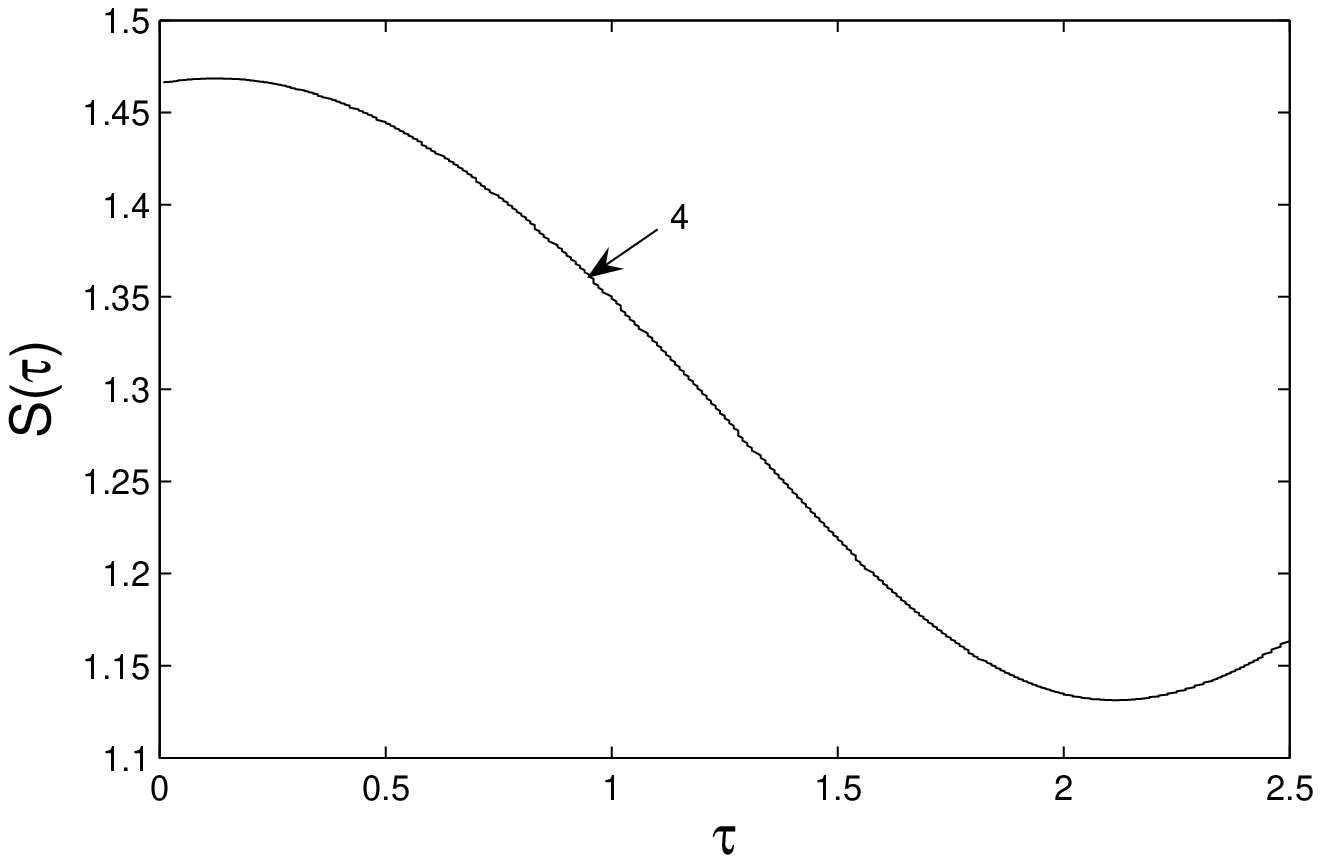}
\caption{(1) is $S(\tau)$ for variables $\psi_1 \& \psi_3$ , (2)
for $\psi_1 \& \psi_2$. Line (1) shows complete synchronization
while the other is not synchronized. Both these lines are
plotted with no phase difference applied (3) and (4) gives the
similarity function for the variables $\psi_1 \& \psi_3$ and
$\psi_1 \& \psi_2$ in the presence of phase difference
$\theta=0.5 \pi.$  The second figure shows line 4 where the dip
can be observed clearly.
} %
\label{similarity}
\end{figure}

\begin{figure}[tbh]
\centering
\includegraphics[width=8cm]{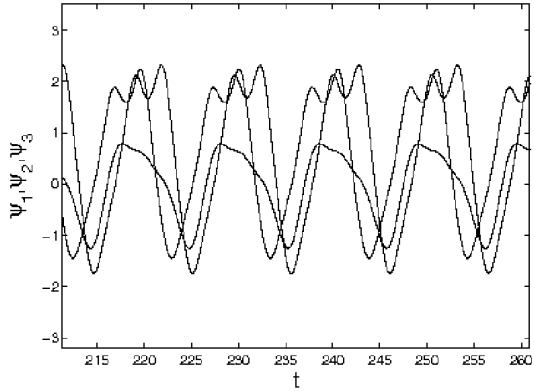}%
\caption{The variables corresponding to the three JJs are ploted against time which indicates the periodic behavior.
} %
\label{period}
\end{figure}

The level of mismatch of chaotic synchronization can be given
quantitatively by taking the similarity function $S(\tau)$ as a
time averaged difference between the variables $\psi_i$ taken
with time shift $\tau$ \cite{kur}
\begin{equation}
S^2(\tau)=\frac{\langle\left[\psi_1(t+\tau)-\psi_2(t)\right]^2\rangle}
{\left[\langle\psi_1^2(t)\rangle\right]\left[\langle\psi_2^2(t)\rangle\right]^{1/2}}
\end{equation}
and
\begin{equation}
S^2(\tau)=\frac{\langle\left[\psi_1(t+\tau)-\psi_3(t)\right]^2\rangle}
{\left[\langle\psi_1^2(t)\rangle\right]\left[\langle\psi_3^2(t)\rangle\right]^{1/2}}.
\end{equation}
and searching for its minimum $\sigma=min_{\tau} S(\tau).$ If
$\psi_1(t)=\psi_3(t),$ then $S(\tau)$  has a minimum value
$\sigma=0$ for $\tau =0$. If both $\psi_1(t)$ and $\psi_3(t)$
are independent then $S(\tau)\approx1$ for  all the time. Line 1
in Fig.~\ref{similarity} shows complete synchronization between
the end junctions  and line 2 shows that the outer and middle
junctions are desynchronized when no phase difference is
present. A minimum of $S(\tau)$ indicates the the existence of a
time shift between the two variables related to the phase shift.
The amplitudes are uncorrelated in this regime, but phase
correlation is present  as indicated by lines 3 and 4 in the
presence of a phase difference between the applied fields.
On the application of a phase difference of $\pi/2$ the dynamics
changes to periodic one as can be observed from
Fig.~\ref{period}
\section{Result and Discussion}
We consider a parallel array of JJs with linking resistor $R_s$
and the conditions for synchronization is discussed. The outer
junctions being symmetric, can possess identical solution and
hence may synchronize depending on the parameter values.  Linear
stability analysis is done to find the stability of the
synchronous solution of the outer junctions. The sum of
conditional Lyapunov exponents calculated for the outer
sub-system is found to be negative indicating stable synchronous
state. From symmetry considerations we show that all three
junctions could be synchronized only in the absence of an
external field. Similarly in an array of N Josephson junctions,
$N/2$ identical solutions may exist if  the number of junctions
is even and $\frac{N+1}{2}$ solutions may exist if the number of
junction is odd.  In the presence of a small phase difference,
the system desynchronizes due to the asymmetry induced by the
phase difference. As the phase difference is increased, in the
case of three junctions all the three junctions act as if they
are driven by different driving fields having the same
frequency, but different phases. A phase synchronization is
observed between all the three junctions and the motion becomes
 periodic. Thus, suppression of chaos can be obtained
in Josephson junction systems in the presence of a phase
difference between the applied fields and this property may find
applications in the working of devices constructed using JJs.

\acknowledgements
The authors acknowledge DRDO, Government of
India for financial assistance through a major research project.


%

\begin{thebibliography}{99}
\bibitem{huber}B. A. Huberman, J. P. Crutchfield and N. H. Packard,
 Appl. Phys. Lett.\textbf{37},  750 (1980).

\bibitem{humier}D. D'Humieres, M. R. Beasley, B. A. Huberman, and A. Libchaber,
\pra \textbf{26}, 3483 (1982).

\bibitem{braiman}Y. Braiman and I. Goldhirsch,
\prl \textbf{66}, 2545 (1991).



\bibitem{fesser}K. Fesser,
A. R. Bishop and P. Kumar, Appl. Phys. Lett.\textbf{43}, 123
(1983).

\bibitem{schieve}W. C. Schieve,  A. R. Bulsara and E. W. Jacobs, \pra \textbf{37},
3541, (1988).

 \bibitem{pozzo}Ezequiel N. Pozzo and Daniel Domínguez, \prl \textbf{98},
057006, (2007).



\bibitem{schul}H. Schulze, R. Behr, F. Müller, and J.
Niemeyer,  Appl. Phys. Lett. \textbf{73}, 996 (1998).

\bibitem{benz}P. Benz and C. A. Hamilton,  Appl. Phys. Lett. \textbf{68}, 3171 (1996).

\bibitem{pecora}L. M. Pecora and T. L. Carroll,
     Phys. Rev. Lett. \textbf{64}, 821 (1990).
\bibitem{pik}A. Pikovsky,M.Rosenblum and J.Kurths,
    Synchronization: A Universal concept in Nonlinear Science, Cambridge University press, Cambridge, 2001.
\bibitem{kur}M. G. Rosenblum, A. S. Pikovsky, and J. Kurths,
       Phys. Rev. Lett. \textbf{78},  4193(1997).

\bibitem{terry} J. R. Terry et.al,
       \pre \textbf{59}, 4036 (1999).

\bibitem{shua} J. W. Shuai and K. W. Wong,
       \pre \textbf{57}, 7002 (1998).


\bibitem{bapt} M. S. Baptista, C. Zhou, and J. Kurths,
      Chinese Phys. Lett. \textbf{23}, 560 (2006).

\bibitem{hal}J. Hale,
 Functional Differential Equations, Springer-Verlag, New York, 1971.
\bibitem{pec}L. M. Pecora and T. L. Carroll,
     Phys. Rev. Lett. \textbf{80}, 2109 (1998).

\bibitem{yin}H.W.Yin, J.H.Dai and H.J. Zhang,
   Phys. Rev. E \textbf{58},  5683 (1998).

\bibitem{kim}K.-T. Kim, M.-S. Kim, Y. Chong and J. Niemeyer,
Appl. Phys. Lett. \textbf{88}, 062501 (2006).


\bibitem{chitra}Chitra R Nayak and V.C. Kuriakose,
      Phys.Lett.A \textbf{365}, 284 (2007).

\bibitem{chevr} D. Chevriaux, R. Khomeriki, and J. Leon,
\prb \textbf{73}, 214516 (2006).


\bibitem{blackburn}J.A.Blackburn, G.L.Baker and H.J.T.Smith,
 \prb \textbf{62},  5931 (2000).


\bibitem{vincent} U.E. Vincent et.al,
Chaos \textbf{14}, 1018 (2004). \\S.L. Ross, Differential
Equations, John Wiley \&
 Sons, Third Edition (1984).

\bibitem{alex} A. S. Landsman and I. B. Schwartz,
\pre \textbf{75}, 026201 (2007).

\end{thebibliography}
\end{document}